\documentstyle[preprint,aps]{revtex}
\input psfig
\begin{document}
\date{Revised version: 6 June 1997}
\title{Dynamics of Viscous Amphiphilic Films
Supported by Elastic Solid Substrates.}
\author{M. V. Voinova$^{1,2}$, M. Jonson$^2$ and B. Kasemo$^2$}
\address{$^1$ Department of Theoretical Physics, Kharkov State University,\\
 Kharkov 310077, Ukraine\\
$^2$ Department of Applied Physics, Chalmers University of Technology and 
G\"oteborg University,\\
S-412 96, G\"oteborg, Sweden}
\maketitle
\begin{abstract}
The dynamics of amphiphilic films deposited on a solid surface is analyzed
for the case when shear oscillations of the solid surface are excited. The
two cases of surface- and bulk shear waves are studied with
the film exposed to gas or to a liquid.
By solving the corresponding  dispersion equation and the wave 
equation while maintaining the 
energy balance we are able to connect the
 surface
density and the shear viscosity of a fluid amphiphilic  
overlayer with experimentally accessible
damping coefficient, phase velocity, dissipation factor and resonant
frequency shifts of shear waves.

PACS number(s): 68.15.+p, 68.60.Bs, 43.35.Bf
\end{abstract}
\pagebreak
\baselineskip=24pt
\section{Introduction.}

Measurements of the properties of thin adsorbed films have since long been an 
important subfield of interface studies \cite{1,2,3,4,5,6,7,8,9,10}. 
Scientific and technological applications of thin films
like Langmuir-Blodgett (LB) films \cite{11,12,13,14,15,16,17,18,19,20}, 
self-assembled monolayers (SAM) \cite{21} and protein monolayers and 
multilayers \cite{22,23} have stimulated detailed studies of the
physical properties of such structures.
With these model systems it is possible to construct multilayer amphiphilic 
films with controlled
monolayer thickness $h_{j}\simeq  25\AA$ and to create different coupling
situations between the adsorbed film and the chemically modified substrate 
\cite{12,13,14,15,16}. 
Specifically measurements of surface density and 
viscosity of these films are important both scientifically and, e.g., 
with respect to their 
possible application in acoustical biosensors in which they can serve as
sensitive elements \cite{13,16,18,19}.

The dynamic (shear) viscosity of amphiphilic films is a very important
rheological characteristic of a protein film or a fluid membrane, which is 
strongly dependent on temperature and phase state of the film, and on
environmental conditions \cite{13,24,25}.   
Usually LB or SAM film acoustic sensors operate in vacuum or in a gaseous
enviroment. In contrast, a lipid-water system or a protein layer
adsorbed onto a solid surface in a fluid experiences 
``wet" conditions or a bulk aqueous medium [23,26] at room temperature, 
which can change the viscosity of the film. In addition, the biomolecular 
layer adsorption process is extremely sensitive to the nature of the solid 
substrate in the sense that it 
can modify the overlayer structure \cite{13,16,20,21,22}. In view of this,
a rigorous description of the dynamics of the existing interfaces of 
combined solid substrate-adsorbed layer-bulk liquid systems, as well as of  
solid surface-amphiphilic film-gas systems seems to be greatly needed.

In the present paper we investigate the dynamics 
 of thin amphiphilic layer(s) attached to a solid substrate
 oscillating in a shear mode and in contact with (i) a gas or
(ii) a bulk newtonian liquid. Figs.~(1-3) show schematically the
geometry of such sandwich systems.

Amphiphilic bilayer films can imitate the behavior of a bilayer lipid
membrane. The constituent bilayer molecules are composed of hydrophilic
groups attached to hydrophobic chains of different length.
The phase diagrams of such amphiphilic molecules demonstrate a rich 
variety of properties and behavior depending on temperature and water 
content [13,20,24,27,28]. Above the liquid crystalline-to-gel transition
the hydrocarbon chains are approximately liquid  
and a bilayer membrane behaves like a
two-dimensional fluid in the lateral plane due to the vanishing
shear modulus of elasticity [27,28]. Such a peculiar fluid is isotropic
in its plane but it is  anisotropic in the normal
direction due to sublayered head-and-tail structure [24].  Viscous layer
response on shear deformation  can be characterized by different shear
viscosity coefficients, namely, the surface viscosity, $\eta_s$ [13], and
a bulk shear viscosity, $\eta_M$ (a component of tensor $\eta_{jklm}$,
index 'M' corresponds to the classification according to  Ref.~28). 

When the system is immersed in water, anisotropy can
arise from water being trapped in the layer. This can result in a value
of the overlayer viscosity  $\eta_M$,
which is distinctly different compared to the case of a gaseous (air) 
enviroment. 
 This is especially
important for protein adsorbed layers with nonuniform interfacial
domain structure and nonuniform water distribution in the normal direction. 
Such nonuniformity effects have been analyzed in a recently published
hydrodynamic model of ``porous" polymer surface film in liquids
(for a review see Ref.~26 and references therein). 

In the present paper we have considered both the surface  viscosity
 $\eta_s$  and  the bulk shear 
viscosity $\eta_M$  of adsorbed amphiphilic films. These 
can be measured in different acoustic experiment,
when acoustic shear waves 
propagate along the plane of the layer (horizontally polarized surface
shear waves) and in the direction normal to it (bulk acoustic shear 
waves). In order to be able to describe both air- and liquid enviromental 
experiments,  corresponding shear viscosity coefficients,
 $\eta_{M}~ 'in ~air'$  and  $\eta_{M}~'in~ liquid'$ have been noted.

The paper is organized as follows: in the framework of continuum mechanics
we derive in Section II the dispersion equation for elastic shear waves
propagating in the system. We also treat the interaction between 
horizontally polarized elastic surface
shear waves (SH SSW) and a fluid layer adsorbed
on the surface of a semi-infinite solid substrate.
In Section III we analyze the response 
of bulk acoustic shear waves (BAW) 
propagating in a finite quartz plate with a viscous overlayer.
Our results allow us to connect 
the surface density and shear viscosity of amphiphilic overlayers with 
experimentally accessible damping coefficient, 
phase velocity, dissipation factor and  resonant frequency shift 
of shear waves. These can be 
measured by modern piezo-acoustical devices of different types
\cite{23,26,29}.     
\vspace{10 mm}

\section{ Dynamics of a Viscous Bilayer Film on a Solid Substrate
Oscillating in a Gaseous Medium or in Vacuum.}

Within continuum mechanics [30,31] the dynamics of 
a liquid film  can be described by the  equation
$$
\rho \frac{dv_{j}}{dt} = \rho \left( \partial v_{j}/\partial t 
+  \left( {\bf v} \nabla \right) v_{j} \right) = \partial_k \sigma_{jk}
\quad .
$$
This corresponds to the Navier-Stokes equation for the motion of a viscous
liquid. Recently  it was shown that hydrodynamical properties of fluid 
amphiphilic (lipid) films can be analyzed within the continuum mechanics 
scheme with viscous tensions defined as follows [28]: 
$$
\begin{array}{c}
\sigma_{jk} = {\cal A}_{lm}\eta_{jklm}\\
{\cal A}_{lm}= \frac{1}{2} \left( \partial_l v_m + \partial_m v_l \right),
\, \partial_l \equiv \partial /\partial x_l \quad .
\end{array}
$$
Here  $\eta_{jklm}$ is a viscosity matrix which were introduced
 explicitly in Ref.~\cite{32}.

For a fluid lipid layer in the incompressible liquid approximation and
in the absence of pressure gradient, these expressions can be simplified 
\cite{28}. In particular, for the case when a strain is applying along
 the $x$ direction, one finds
$$
\begin{array}{l}
\sigma_{xy} = \eta_{M} \left(  \partial_y v_x\right).\\
\end{array}
$$
Here the viscosity coefficient $\eta_{M}$ is
a component of the viscosity matrix $\eta_{jklm}$; the
 $y$-axis is perpendicular to the layer plane. 

In our model we use a notation  $\eta_{M}~ (liquid)$  and
 $\eta_{M}~(gas)$ for describing fluid films bounded (in different
 experiments) by a liquid or a gas.

Below we analyze the influence of viscosity and surface density of 
a thin bilayer on the phase velocity and the damping of elastic 
shear waves.

\subsection*{Surface Shear Wave Propagation: The Semi-infinite
Quartz Crystal Substrate}

We treat here horizontally polarized surface elastic waves on a solid 
substrate interacting
 with a thin fluid double layer, which on the other side has
an interface to vacuum or to a gaseous phase (Fig.1).
Surface waves on plane interfaces have an amplitude which decays exponentially
with normal distance from the solid surface on which they propagate [2]. 
Phase velocities of surface
 shear waves (SSW) $V$ are lower than those of bulk shear waves  
$V_{0}$ in the semi-infinite elastic half-space (i.e. the substrate). 

Let us now consider how the shear vibration of the substrate generates
a viscous wave in the adjacent fluid layer \cite{2,5,17}.  
In the framework of fluid mechanics, the motion of an adsorbed viscous film is 
described by the linearized Navier-Stokes equation:
\begin{equation}
\label{1}
\begin{array}{l}
\partial v_{x} \left( y,z,t) \right) /\partial t = \nu_1 \Delta v_x, \\
\nu_1 \equiv \eta_1/\rho_1, \\ 
\eta_1 = \eta_{M} ~(gas).
\end{array}
\end{equation}
 Equation (1) is valid in the regime of
small Reynolds numbers. 
$$
R = \frac{\omega u_0 h_1}{\nu_1} \quad \ll 1,
$$
for small oscillation frequency $\omega$ and -amplitude $u_0$ and for a thin 
overlayer thickness $h_1$.
Here $v_x$ denotes the $x$-component of the velocity of the fluid film
and the $y$-axis is perpendicular to the $z$-direction of wave propagation. 

The boundary conditions at the fluid-solid interface ($y=0$) correspond to the
assumption of {\sl no slip} conditions [6,26,28,30]
\begin{equation}
\label{2}
\begin{array}{r}
y=0: \qquad
v_x = \frac {\partial u_x \left( y,z,t \right) }{\partial t}, \\
\sigma_{yx} = \eta_1\frac{\partial v_x}{\partial y},\\
\sigma_{yx} = C_{44}\frac{\partial u_x}{\partial y}.
\end{array}
\end{equation}
Here $u_x$ is a component of the substrate boundary displacement vector,
$$
u_x = u_0 {\rm exp}(\kappa_t y + iqz){\rm exp}(i\omega t),
$$
and $\kappa_t = {(q^{2} - \rho_0 {\omega}^2/C_{44})}^{1/2}$ is the inverse
penetration depth of surface shear waves into the solid substrate; $\omega$,
q are wave frequency and wave number, while $\rho_0$ and $C_{44}$ are the
substrate density and shear modulus, respectively. The substrate
boundary motion is described by the equation 
$$
\rho_0 \ddot{u_x} = C_{44} (\frac{\partial^2 u_x}{\partial y^2} +
\frac{\partial^2 u_x}{\partial z^2}),
$$
which corresponds to elastic vibrations of the solid surface.

Boundary conditions at the moving interface between fluid 1 and fluid 2  
($y = h_1$, Fig.~1) follow from the condition that the friction forces must be 
the same \cite{30}:
\begin{equation}
\label{3}
\begin{array}{r}
y = h_1,\qquad n_k \sigma_{ik}^{(1)} = n_k \sigma_{ik}^{(2)},\\
\sigma_{ik}^{(1)} = -p_1 \delta_{ik} + \eta_1(\frac{\partial {v_i}^{(1)}}
{\partial x_k} + \frac{\partial {v_k}^{(1)}}{\partial x_i}),\\
\sigma_{ik}^{(2)} = -p_2 \delta_{ik} + \eta_2(\frac{\partial {v_i}^{(2)}}
{\partial x_k} + \frac{\partial {v_k}^{(2)}}{\partial x_i}),\\
v_x^{(1)} = v_x^{(2)}, \qquad v_z^{(1)} = v_z^{(2)} = 0,
\end{array}
\end{equation}
At the free surface, finally, the boundary condition is:
\begin{equation}
\label{4}
y = h_2, \qquad \eta_2 \frac{\partial v_x^{(2)}}{\partial y} = 0,
\end{equation}

Equations (1-4) together with the equation for the elastic substrate motion
lead to the following dispersion equation for 
horizontally polarized surface shear waves in our system:
\begin{equation}
\label{5}
\kappa_t = \frac{i\omega\eta_1\xi_1}{C_{44}} \frac{(\frac{\xi_1}{\xi_2}
+ \epsilon \tanh(\Delta h\xi_2))exp(2h\xi_1) - (\frac{\xi_1}{\xi_2} 
- \epsilon \tanh(\Delta h\xi_2))}{(\frac{\xi_1}{\xi_2} + \epsilon 
\tanh(\Delta h\xi_2))exp(2h\xi_1) + (\frac{\xi_1}{\xi_2} - \epsilon 
\tanh(\Delta h\xi_2))} ,
\end{equation}
where $\xi_j = (q^2 + i \omega/\nu_j)^{1/2},\quad j = 1,2$,
$\epsilon \equiv \eta_2/\eta_1$, and $\Delta h = h_2 - h_1$.
 
Here we introduce the viscous penetration depth
 $\delta \equiv {(2\nu/\omega)}^{1/2}$
corresponding to the distance over which the
transverse wave amplitude  falls off by factor of $e$.
For two sufficiently thin viscous overlayers one can assume 
that $h_j/\delta_j\ll 1$, and we will always consider the long wavelength
limit, where $q^{2} \ll \delta^{-2}$.
Within these limits, we can from Eqn.~(5) find the SSW
 damping coefficient as a function of the viscosities and the surface
 densities of the overlayers.

The result for the damping coefficient $\Gamma$ of the SSW --- simply the 
imaginary part of the wave vector $q$ ---  is
\begin{equation}
\Gamma = \Gamma_{mon} \left\{1+ \frac{\eta_{2s}}
{\eta_{1s}}\left( 1+\frac{\eta_{1s} \rho_{2s}}{\eta_{2s} \rho_{1s}}
\left( 1+\frac{\eta_{2s}}{\eta_{1s}}\right)\right)\right\}. \,
\end{equation}
Here $\eta_{js} \equiv \eta_{M} h_{j}$ is the surface viscosity 
component of each layer, $\rho_{js} \equiv \rho_{j} h_{j}$ is a
corresponding surface density, 
and $\Gamma_{mon}$ denotes the damping coefficient for the case of a 
monolayer \cite{17},
$$
\Gamma_{mon}=\frac{q_{0}\omega^{3}\rho_{1s}\eta_{1s}}{C_{44}^{2}}  . 
$$
One can estimate the value of $\Gamma$ to be $\sim 10^{-5}$ for a 
phospholipid bilayer. 
Experimental data for egg lecithin at $T=25^\circ C$  are taken from Ref.~\cite{24}, which provides the following
values:
bilayer thickness
$h_{b}= 46\AA$,  bilayer density $ \rho\approx 1$~g/cm$^3$,  bilayer
``microviscosity" (measured by a probe technique) 
$\eta\approx 1.2$~dyn$\cdot$s/cm$^2$. The resonator frequency is
 $\omega_{0}\approx 10^9$Hz and $C_{44}\approx 2.6$ dyn/cm$^2$ is 
the corresponding quartz shear modulus.

Using the dispersion equation (5) one can also obtain the change in
the SSW phase
velocity, $\Delta V/V_0$, caused by the presence of the adsorbed 
bilayer. One finds that
\begin{equation}
\frac{\Delta V}{V_0} \approx \frac{1}{2}\bigl(\frac{\omega \rho_{s_1}}
{V_0 \rho_0}\bigr)^2 \bigl(1+\frac{\rho_{s_2}}{\rho_{s_1}}\bigr)^2
\end{equation}
$$
V_0 \equiv \sqrt{C_{44}/\rho_0} = const  .
$$

As a concequence of the no slip- and thin layer assumptions,
 the SSW velocity 
shift is sensitive only to the overlayer surface densities $\rho_{s_j}$ and
 not to their viscosities. This corresponds to the Love
 type of wave propagation.
Equation (7) allows one to determine the surface density of
the upper or lower half of the
bilayer if the density of the other layer is known or can be
determined by an independent
 experiment. For the lipid bilayer (with the same parameters
 as above) the SSW phase velocity shift is
 small, $\Delta V/V_{0}\sim 10^{-8}$. However, this shift
 may be detectable (of order $\sim 10^{-6}$) at high frequences,
 $\omega \approx 2\pi 10^9$Hz, since the SSW velocity change 
is a quadratic function of frequency. Such high frequencies can be generated
by modern acoustic SSW devices \cite{29}.
At low frequences, $f_\sim^< 10^{8}$Hz, it is more reasonable
 to use another type of shear waves for surface density analysis  
excited by quartz crystals oscillating in the thickness shear mode
(see Section III.B below).

\section{Solid substrate with viscous overlayer 
oscillating in a bulk viscous medium.} 

\subsection*{Resonant Frequency and Dissipation Factor for Bulk Shear Waves:
The Finite Quartz-Crystal Resonator}

In addition to ``genuine" surface waves, another type of shear
 waves can propagate in the plane of the
overlayer-substrate interface. These are
 bulk acoustic waves (BAW). 
In contrast to the SSW case considered above, the acoustical response of
unloaded BAW resonators depends on the quartz slab 
thickness $h_0$ as well as on its density $\rho_0$ and shear
modulus $C_{66}$
$$          
f_0 = (C_{66}/\rho_{0})^{1/2}h_{0}
$$  

It has been shown  \cite{3,6,23,26,33} that the resonant frequency
of such quartz plates  decreases when its surface
is coated with
an overlayer constituting a mass load. 
For a thin surface film, uniformly covering the entire vibrating area of
the slab, Sauerbrey deduced a linear relation 
between the frequency change and the added mass per unit area in a
vacuum (gaseous) environment \cite{3}.
Kanazawa and Gordon \cite{6} have later described the resonant response of
quartz oscillators in bulk viscous liquids. Recent experimental
 and theoretical investigations of the
BAW in composite media (see Refs.~23, 26, 33 for a review) have demonstrated
possible applications to the study of nonuniform polymer and protein thin films
and have included both resonant frequency and viscous dissipation
in the interface region. Experimentally, it is convenient to obtain
this viscous dissipation as the width of the resonant frequency [26]
or as the dissipation factor $D$ [23]
$$
D = E_{dissipated}/2\pi E_{stored} = 1/\pi f\tau
$$
where $\tau$ is the time constant for the decay of the vibration amplitude.

In this Section we calculate the shifts $\Delta f$ of the resonance frequency
and  $\Delta D$ of the
dissipation factor, respectively, for the case of a viscous amphiphilic 
(lipid) film attached without slip to the surface of the quartz slab
 oscillating 
in the thickness shear mode in the presence of
a bulk viscous liquid. 

The wave equation for shear waves propagating in the vertical direction
(Fig.2) is
\begin{equation} \label {W}
\frac{\partial^2 u_x(y,t)}{\partial y^2}=
\frac{i\rho_1 \omega}{\eta_1}u_x(y,t), \qquad \eta_1 \equiv \eta_{M}~(liquid) \,.
\end{equation}
Its general solution has the form
\begin{equation} 
u_x = e^{i\omega t}\left( U_1 e^{-\xi_1 y} + U_2 e^{\xi_1 y} \right)
\end{equation}
$$
\xi_1 = \left( 1 + i \right)/\delta_1\,
$$
Using the same boundary conditions (2-4) as above, we find for the
$x$-component of the liquid velocity:
\begin{equation} \label {S}
v_x=v_0\frac{e^{\xi_{1} y}+Ae^{-\xi_{1} (y-2h_{1})}}{1+Ae^{2h_{1}\xi_{1}}}
\end{equation}
$$
A= \frac{\delta_2 + \epsilon\delta_1 \tanh \Biggl[(1+i)\Delta h/\delta_2 
\Biggr]}{\delta_2 - \epsilon\delta_1 \tanh \Biggl[(1+i)\Delta h/\delta_2 
\Biggr]} 
$$

The shift of resonant frequency $\Delta f$ and  dissipation factor
$\Delta D$, due to the overlayer on the substrate, can be calculated from the
balance between dissipated and stored energy in the system \cite{33,26,23}.
Using this energy balance, we find from Eq.~(11):

\begin{equation}
\Delta f \approx - Im\left(\eta_{1} \xi_1 \frac{A e^{2h_{1}\xi_1} - 1}
 {A e^{2h_{1}\xi_1}+1}\right)/2 \pi \rho_{0} h_{0}\\
\end{equation}
\begin{equation}
\Delta D \approx  -Re\left(\eta_{1} \xi_1 \frac{A e^{2h_{1}\xi_1} -1}
{A e^{2h_{1}\xi_1}+1}\right)/\pi f \rho_{0} h_{0}\\
\end{equation}

In the limiting case $h_1 /\delta_{1}\ll 1$, $ \Delta h /\delta_2 \gg 1$
 we obtain  for the resonant frequency shift
\begin{equation} \label {D}
\Delta f_{res}= -f_0 \Biggl(\frac{f\eta_2\rho_2}{\pi \rho_0 C_{66}}\Biggr)^{1/2}
\Biggl\{ 1+h_1\rho_1\Biggl(\frac{4\pi f}{\eta_2\rho_2}\Biggr)^{1/2}-
(4\pi f \eta_2 \rho_2)^{1/2}\frac{h_1}{\eta_1} \Biggr\}
\end{equation}
This expression contains contributions from both the lower and the upper
layer. For sufficiently viscous solutions,
 $\eta_{2}\approx \eta_{1}\rho_{1}/\rho_{2}$,
 the second and third terms 
may be of the same order.

In contrast, the result for $\Delta D$ is practically insensitive to the 
ultrathin
 overlayer parameters. In the linear approximation, corresponding to a small
value of $h_{1}/\delta_{1}$, it depends only on the upper (bulk)
 liquid viscosity and its density:
\begin{equation} \label {F}
\Delta D = 2f_0\Biggl( \frac{\eta_2\rho_2}{\pi f\rho_0 C_{66}}\Biggr)^{1/2}
\Biggl\{ 1-\frac{2\pi h_1^2 \rho_1 f}{\eta_1}
\Biggl( \frac{\eta_2\rho_2}{\eta_1\rho_1}- 1\Biggr)\Biggr\}
\end{equation}
In the limit of $\eta_{2}\rightarrow 0$ we obtain the Sauerbrey formula:
\begin{equation}
\Delta f = - \frac{2f_{0}^{2}\rho h}{\sqrt{\rho_{0}C_{66}}}.
\end{equation}

In the opposite case and for $h_1 = 0$, the results of Kanazawa and 
Gordon follow: 
\begin{equation}
\Delta f_{res} = - f_0 \left( \frac{f\eta_2 \rho_2}{\pi \rho_0 C_{66}}
\right)^{1/2}.
\end{equation}

Formulae (14) and (15) form a set of equations.
From these expressions it is possible to determine the $\eta_{M}$ component
 of the overlayer shear viscosity 
 and the thickness (mass) of the adsorbed layer and compare them 
with the results of
surface viscosity measurements. Also, it is possible to determine
  the viscosity  and density of the adjacent bulk liquid
 if parameters of the overlayer are known.

 It is essential to note that our model is valid 
for an arbitrary ratio $\epsilon $ between the 
viscosities $\eta_2$ and $\eta_1$, i.e., even for $\epsilon \ge 1$. 
The latter possibility may be
relevant for a thin amphiphilic substrate-adjacent film under a thick protein
layer (For example, a sandwich structure such as the one schematically shown 
in Fig.4. can be realised by an adsorbed protein layer attached to the 
substrate via an amphiphilic self-assembling monolayer [22]).
\vspace {10 mm}

 \section{  Discussion.}

In addition to its importance in technological applications, the study of 
the hydrodynamic modes
 of fluid amphiphilic films and adsorbed proteins are generally 
important for understanding 
the dynamics of the swelling of lipid-water systems [28,34,35] and the
rheology of biological membranes. It is known, for instance, that living cell
 membranes, e.g. in red blood cells, experience shear stress
 in hydrodynamic flow through capillaries \cite{36,37}. 
As the lipid bilayer matrix
 of the membrane is a two-dimensional incompressible 
liquid adjacent to an elastic protein (spectrin-actin)
 network, the hydrodynamics of this layered structure is governed 
by the coupling of the fluid membrane to shear flows
in an external bulk liquid \cite{36,37}.

The surface viscosity of a fluid amphiphilic film is often considered 
to be a two-dimensional
 analogue of the bulk viscosity and defined as the
coefficient of proportionality between the tangential force per 
unit length
and the gradient in the flow velocity of the liquid. 
These averaged surface characteristics
 can be measured, e.g. by an
interface shear rheometer at air-liquid interfaces \cite{13,25}
or by using the oscillating barrier method in a LB trough [25].
In contrast to $\eta_s$, the ``microscopic" shear 
viscosity  may be defined 
 as a contribution of fluid disordered chains.
The ``microviscosity"  of the
membrane core can be determined experimentally,
e.g. by a probe technique, and describes the local viscosity 
near the probe \cite{24}. This
``microviscosity" is often identified with the true membrane 
viscosity. However,
this experimental technique could give rather different 
results with a strong
dependence on the 
probe material and on the interaction between the probe and its local
surroundings \cite{24}. Both of these viscosities may be changed
after the deposition of an amphiphilic layer on the solid surface.

The results of  our theory can be applied for direct acoustic 
measurements of surface densities
 and/or   
shear viscosities $\eta_{s}$  and $\eta_{M}$ for a
thin amphiphilic film after the transferring onto the solid substrate and
in both the liquid and gaseous experimental conditions.
Another help is a possible strong coupling between the end groups of
lipid layer, LB film or SAM and the surface of the substrate [12-16],
thus for the interfacial solid-fluid region, no slip assumption
seems to be valid. 

In our work we found the dispersion equation
for surface shear acoustic waves with horizontal polarization
and solved it  as well
as the wave equation for shear bulk acoustic waves. The solution
of these equations together with energy balance, allows us to calculate
the analytical expressions for damping coefficients, phase velocities,
dissipation factor and resonant frequency shifts of both types of shear
waves as functions of lipid film surface density and two different 
components of its in-plane viscosity. These acoustic waves can be excited 
by means of two different types of piezoelectric oscillators [23,29].   

We can suggest to measure the acoustical responce of the above-mentioned
sandwich system in the region of lipid phase transition when the
viscosity of the overlayer subject to dramatic changes. For instance,
 during the liquid crystalline to gel transformation of a lipid
membrane, the viscosity of the bilayer changes by more than 10 times
 while the membrane
density remains practically constant [17,24]. It is an interesting 
experimental fact [34]
that the two halves of a bilayer lipid membrane are so weakly coupled,
that they can udergo the thermotropic phase transition independently.
In the accordance with our results, it must provide the changes in
SSW damping coefficient, while the corresponding SSW phase velocity
shift due to the presence of bilayer will be constant.

On a more speculative note, our results may also be valid
 within the segregated (sublayer) structure model of adsorbed protein layers 
(Fig.~3).
In accordance with this model (see, e.g. [21,22]), the structure
of the protein overlayer resembles a surfactant film composed of segregated
head-and-tail regions. Recent neutron scattering experiments \cite{22} 
have given evidence for such an anisotropic sublayer 
structure of protein monolayers
adsorbed onto solid surfaces through the self-assembling amphiphilic
 sublayer.
 In this case, the relation between the upper
 and lower layer viscosities may vary in strong dependence of the
phase state of protein and amphiphilic monolayer, respectively.

\vspace{10 mm}

\section{ Acknowledgements.}

One of us (MV) is grateful to Prof. S .F. Edwards for the 
possibility of visiting the
Polymer and Colloid Group at the Cavendish Laboratory.
We  gratefully acknowledge detailed discussions 
about quartz crystal microbalance experiments with M. Rodahl and A. Krozer 
(CTH/GU) and helpful correspondence concerning recent neutron scattering data
on adsorbed proteins with A. R. Rennie (Cambridge).

This work was supported by the Royal Swedish Academy of Sciences (KVA).

\begin{figure}
\centerline{\psfig{figure=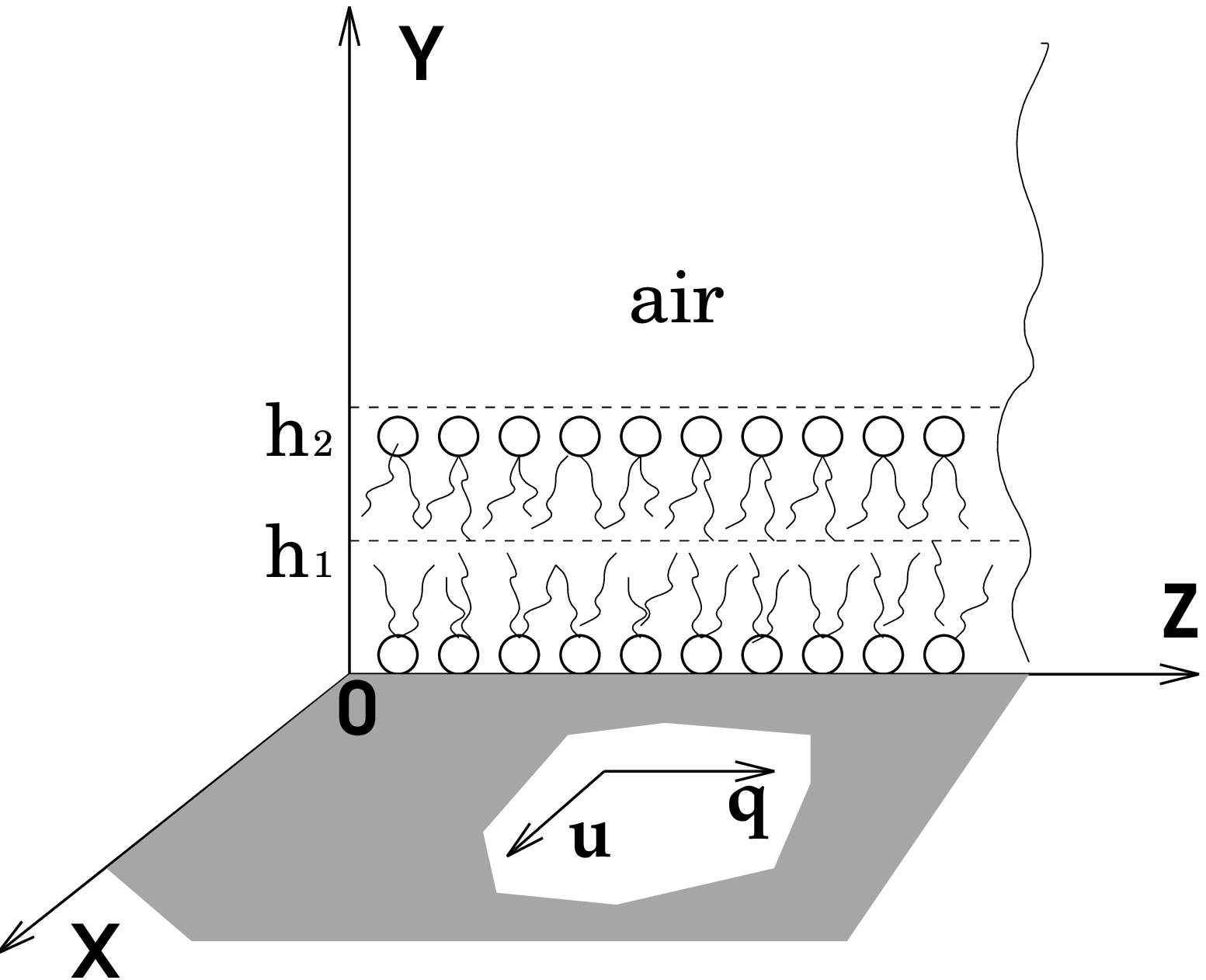}}
\vspace{1cm}
\caption{ Illustration of surface shear waves (SSW) with horizontal
polarization propagating in a system of an elastic half-space with
an adsorbed thin amphiphilic fluid bilayer on top. The system is
in a gaseous enviroment.}

\label{fig1}
\end{figure}

\begin{figure}
\centerline{\psfig{figure=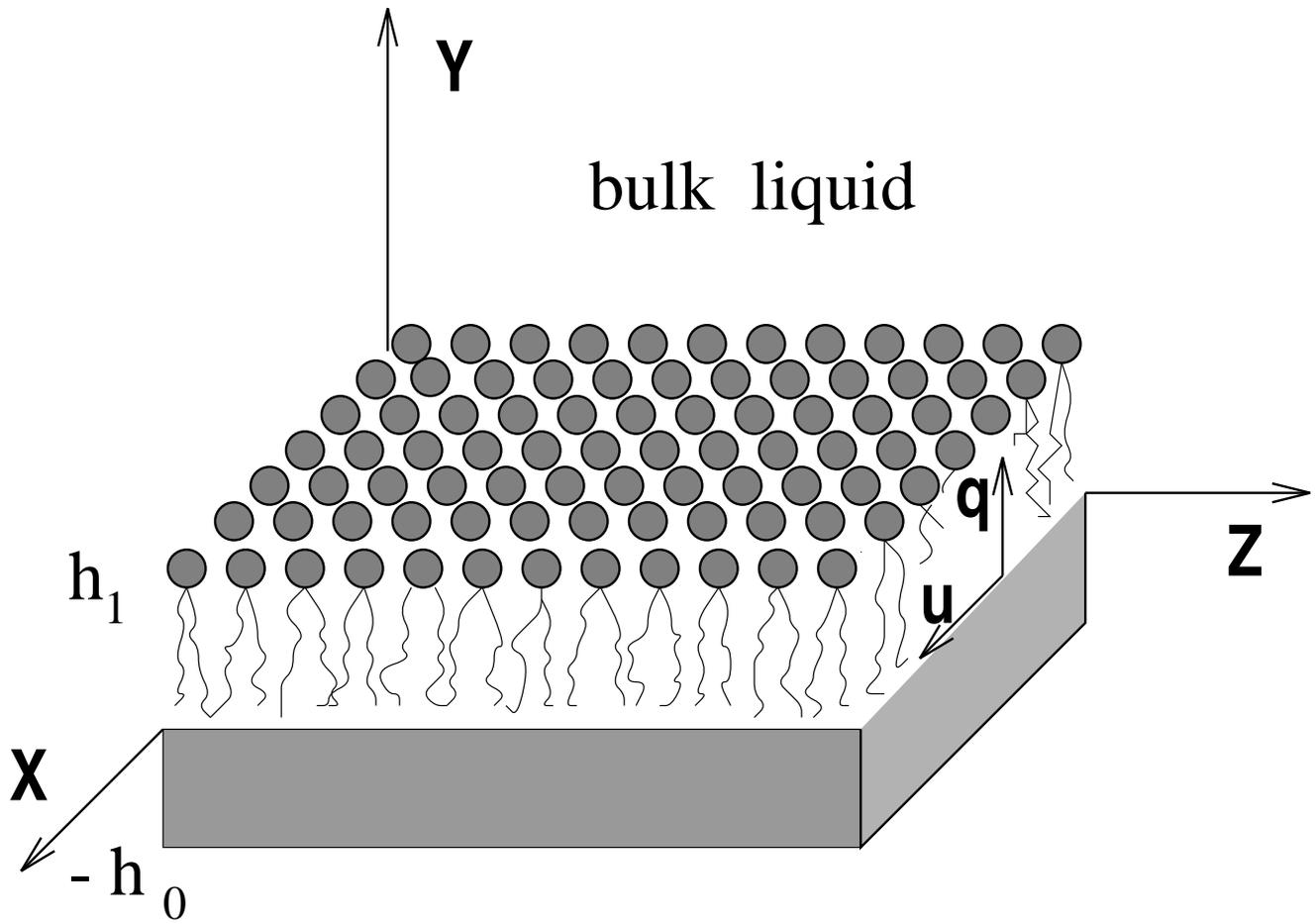}}
\vspace{1cm}
\caption{ shear bulk waves propagating in an
AT-cut quartz plate covered with a thin amphiphilic layer.
The system is immersed in a bulk liquid.}

\label{fig2}
\end{figure}

\begin{figure}
\centerline{\psfig{figure=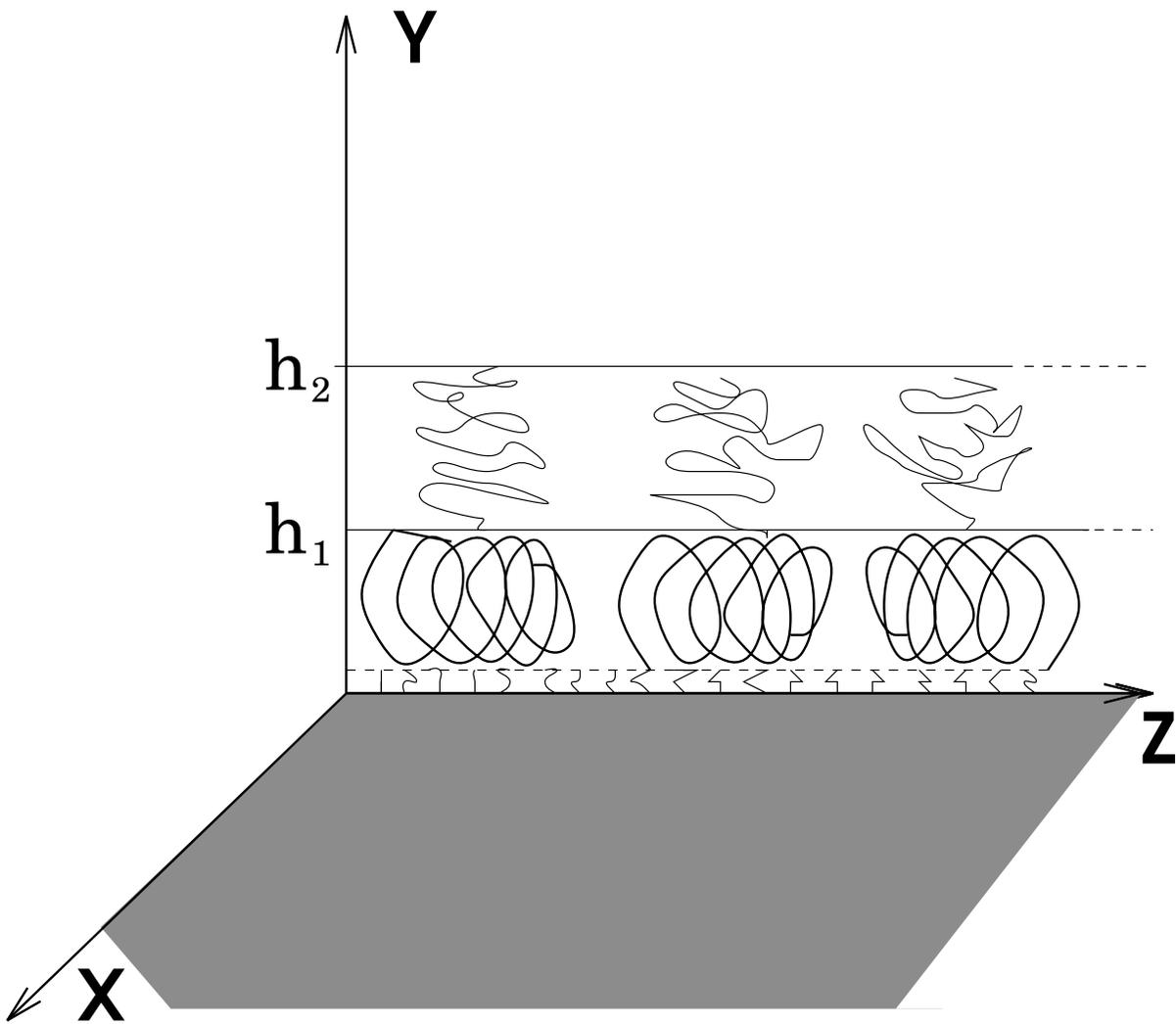}}
\vspace{1cm}
\caption{ Schematic depiction of two distinct
regions of protein layer absorbed from 
bulk solution onto a solid substrate. The dotted line correspondes to
the surface-adjacent amphiphilic layer.}
           
\label{fig3}
\end{figure}

\end{document}